\documentclass[]{spie}  

\usepackage{amsmath,amsfonts,amssymb}
\usepackage{graphicx}
\usepackage[colorlinks=true, allcolors=blue]{hyperref}

\title{Imaging Protoplanets: Observing Transition Disks with Non-Redundant Masking}

\author[a]{Steph Sallum}
\author[a]{Josh Eisner}
\author[a]{Laird M. Close}
\author[a]{Philip M. Hinz}
\author[b]{Katherine B. Follette}
\author[a]{Kaitlin Kratter}
\author[c]{Andrew J. Skemer}
\author[b]{Vanessa P. Bailey}
\author[d]{Runa Briguglio}
\author[e]{Denis Defrere}
\author[b]{Bruce A. Macintosh}
\author[a]{Jared R. Males}
\author[a]{Katie M. Morzinski}
\author[d]{Alfio T. Puglisi}
\author[f]{Timothy J. Rodigas}
\author[a]{Eckhart Spalding}
\author[g]{Peter G. Tuthill}
\author[a]{Amali Vaz}
\author[f]{Alycia Weinberger}
\author[d]{Marco Xomperio}

\affil[a]{Astronomy Department, University of Arizona, 933 North Cherry Avenue, Tucson, AZ 85721, USA}
\affil[b]{Kavli Institute for Particle Astrophysics and Cosmology, Stanford University, Stanford, California 94305, USA}
\affil[c]{Department of Astronomy and Astrophysics, University of California, Santa Cruz, 1156 High Street, Santa Cruz, CA 95064, USA}
\affil[d]{INAF-Osservatorio Astrofisico di Arcetri, I-50125 Firenze, Italy}
\affil[e]{STAR Institute, Universit\'e de Li\`ege, 17 All\'ee du Six Ao\^ut, B-4000 Sart Tilman, Belgium}
\affil[f]{Carnegie Institution DTM, 5241 Broad Branch Rd, Washington, DC 20015, USA}

\authorinfo{Further author information: (Send correspondence to S.S.)\\S.S.: E-mail: ssallum@email.arizona.edu, Telephone: +1 520 621 2288}
\begin{document} 
\maketitle

\begin{abstract}
Transition disks, protoplanetary disks with inner clearings, are promising objects in which to directly image forming planets. The high contrast imaging technique of non-redundant masking is well posed to detect planetary mass companions at several to tens of AU in nearby transition disks. We present non-redundant masking observations of the T Cha and LkCa 15 transition disks, both of which host posited sub-stellar mass companions. However, due to a loss of information intrinsic to the technique, observations of extended sources (e.g. scattered light from disks) can be misinterpreted as moving companions. We discuss tests to distinguish between these two scenarios, with applications to the T Cha and LkCa 15 observations. We argue that a static, forward-scattering disk can explain the T Cha data, while LkCa 15 is best explained by multiple orbiting companions.

\end{abstract}


\section{INTRODUCTION}

Direct images of protoplanets will provide unique insight into the planet formation process. Transition disks, protoplanetary disks with solar system sized inner dust clearings\cite{1989AJ.....97.1451S,2009ApJ...704..496B} present an opportunity to image protoplanets directly. Both observations\cite{2008ApJ...678L..59I,2012ApJ...745....5K,2013ApJ...775...30I,2015Natur.527..342S} and simulations\cite{1986ApJ...309..846L,1999ApJ...514..344B,2006Icar..181..587C} suggest that dynamical interactions with unseen companions may play an important role in creating the clearings. However, stellar mass companions have been ruled out in approximately half of known transition disks\cite{2011ApJ...731....8K,2012ApJ...744..120E}. This leaves the possibility that planetary mass companions are responsible for the disk clearings. 

Imaging protoplanets in transition disks requires both high contrast and high resolution. Compared to mature exoplanets, accreting planets are bright relative to their host stars in the infrared\cite{2015ApJ...803L...4E,2015ApJ...799...16Z}, having contrasts of $\sim 10^{-2} - 10^{-4}$. At the distance of the nearest transition disks ($\sim100$ pc) 10 AU subtends $0.1''$, approximately the diffraction limit of an 8 m telescope at L band ($0.095''$). Searching for planets at $\sim$ AU to tens of AU separations therefore requires the ability to image at high contrast at or within the diffraction limit. 

While traditional direct imaging techniques can probe angular scales of a few $\lambda/D$, the high contrast imaging technique of non-redundant masking (NRM)\cite{2000SPIE.4006..491T} provides much better point spread function characterization, probing scales even within $\lambda/D$. NRM transforms a filled aperture into a sparse array through the use of a pupil plane mask. The Fourier-transformed images yield complex visibilities, which give the amplitude and phase of each baseline in the mask. From the complex visibilities we calculate squared visibilities, the power on each baseline, and closure phases, sums of phases around baselines forming a triangle. Closure phases eliminate atmospheric phase errors; each closure phase is the sum of the intrinsic source phases. Measuring the sums of source phases in this way means that not all source phase information can be recovered from observations. Due to this missing phase information, to understand the source brightness distribution we use both model fitting and image reconstruction. 

Here we present two NRM transition disk case studies, T Cha and LkCa 15. Huelamo et al. 2011\cite{2011A&A...528L...7H} detected a companion candidate in T Cha using 2010 Very Large Telescope (VLT) / NaCo\cite{2003SPIE.4839..140R} masking observations. The best fit single companion model had a separation of 62 miliarcseconds (mas) at a position angle of 78$^\circ$ and a contrast of 5.1 magnitudes at L$'$\cite{2011A&A...528L...7H}. Keck masking observations of LkCa 15 between 2009 and 2010 revealed three infrared point sources within the disk clearing\cite{2012ApJ...745....5K} with separations between $\sim 70 - 100$ mas and contrasts between $\sim 5 - 6$ magnitudes at K and L bands. For both of these objects, alternative scenarios have been discussed where the point source detections are caused not by orbiting companions, but by forward scattered light from the circumstellar disk.\cite{2013A&A...552A...4O,2015MNRAS.450L...1C,2014IAUS..299..199I,2015ApJ...801...85S} We discuss tests to distinguish between companion and scattering scenarios, with applications to our observations of both LkCa 15 and T Cha.

\section{T 	Cha}
\subsection{Previously Published Observations}

In Sallum et al. 2015a\cite{2015ApJ...801...85S} we presented re-reduced archival NRM observations of T Cha taken using VLT/NaCo at L$'$ and Ks bands between 2010 and 2013. We also presented new Magellan/MagAO 2013 L$'$ observations. We fitted single companion models to these datasets and found that the best fit position angle changed between the initial companion candidate detection in 2010 and followup datasets through 2013. 

\begin{figure} [ht]
\begin{center}
\includegraphics[height = 10cm]{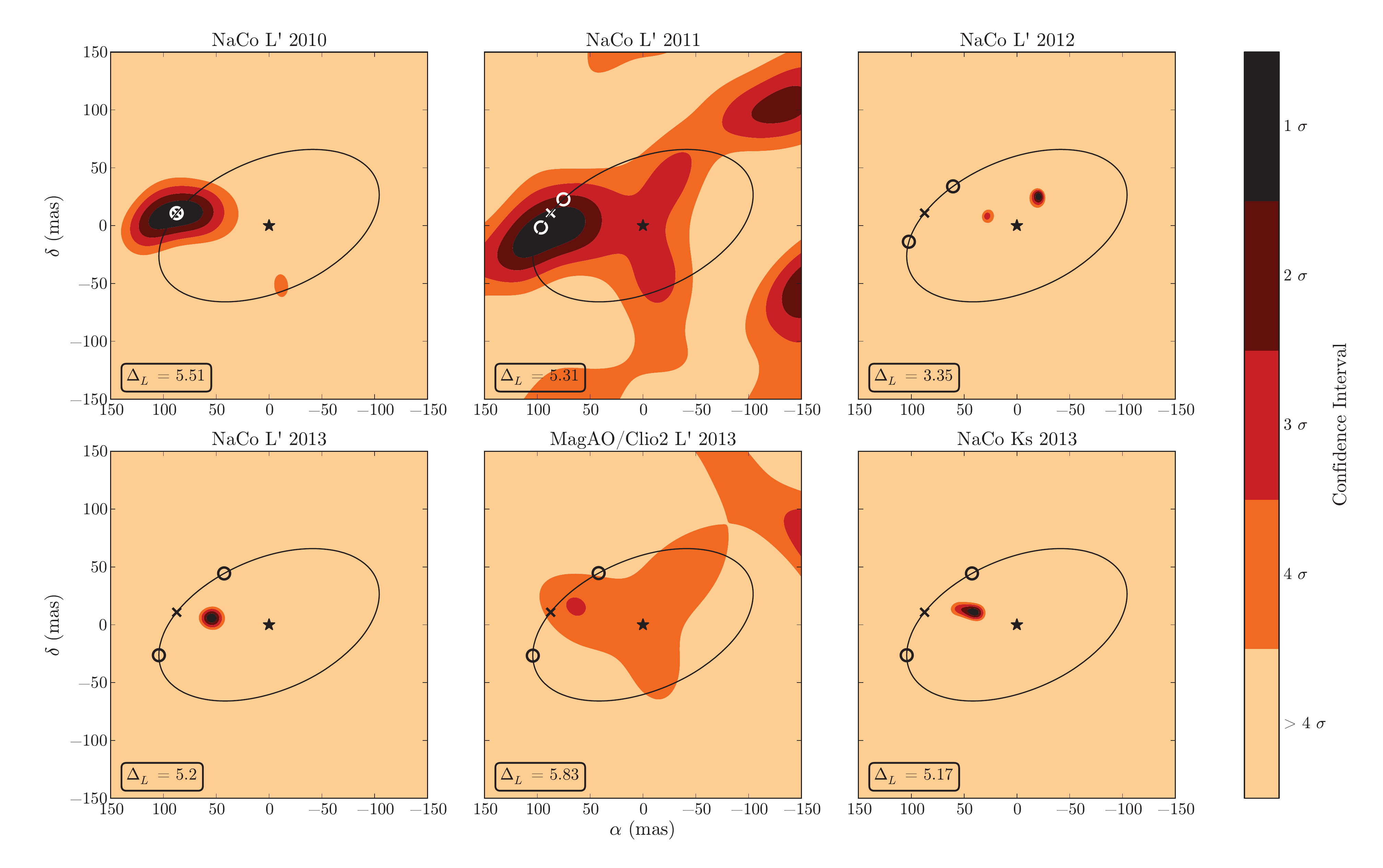}
\end{center}
\caption[tcha_compfits] 
{\label{fig:tcha_compfits} 
From Sallum et al. 2015a. The individual panels show $\chi^2$ slices in right ascension and declination at the best fit contrast ratio for single companion model fits to the T Cha observations. The solid ellipses trace out circular orbits in the plane of the outer disk at the separation of the 2010 best fit companion. In each panel, the x shows the 2010 best fit position, and the circles show predicted positions for orbiting companions at the time of the observations.}
\end{figure}

\begin{figure} [ht]
\begin{center}
\includegraphics[height = 10cm]{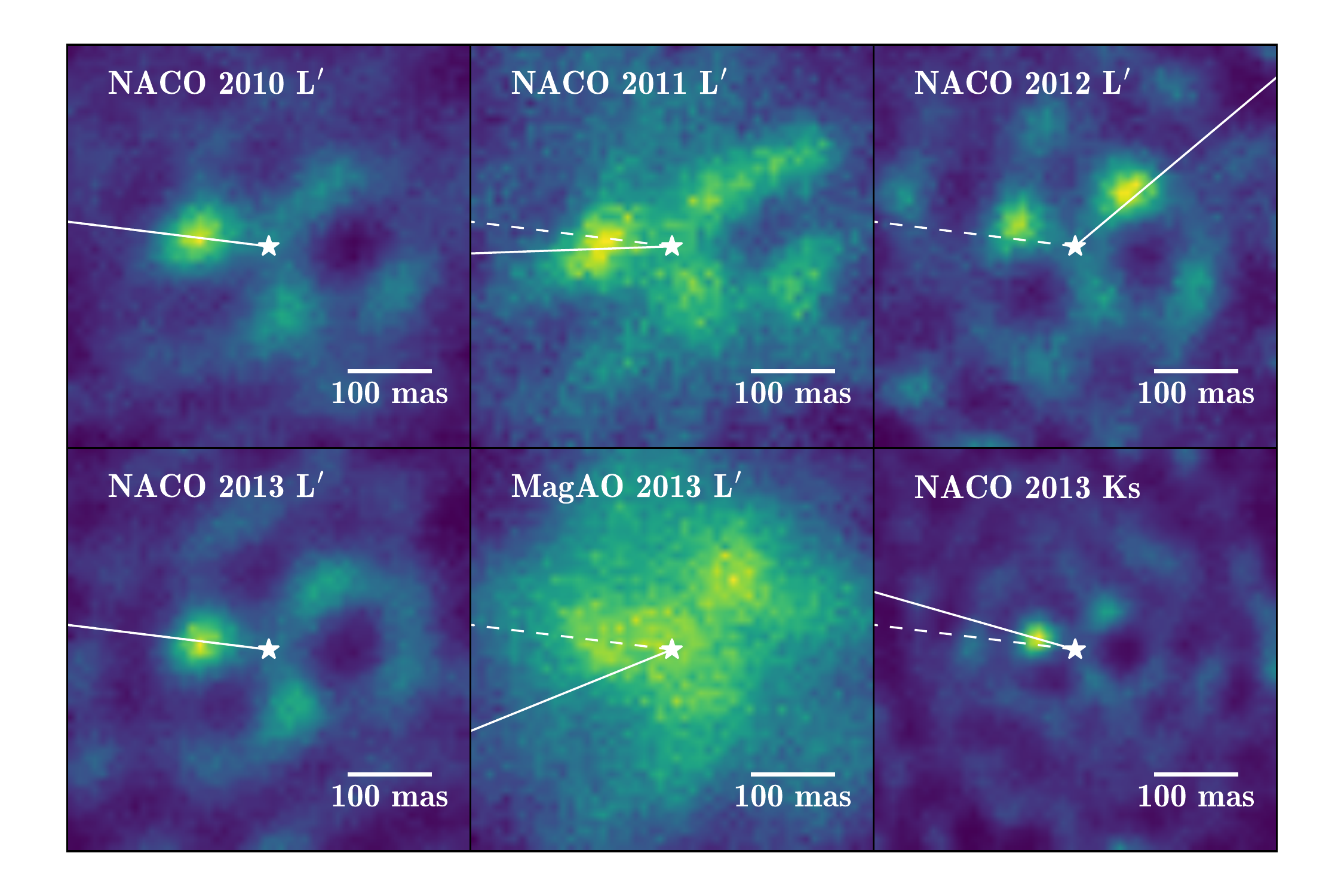}
\end{center}
\caption[tcha_ims] 
{\label{fig:tcha_ims} 
Each panel shows the reconstructed image for a single epoch of T Cha observations in Sallum et al. 2015a. The solid lines mark the best fit single companion position angles, while the dashed lines show the best fit single companion position angle from the 2010 dataset.}
\end{figure} 

\subsection{Companion Scenario}

Figure \ref{fig:tcha_compfits} shows $\chi^2$ slices in right ascension and declination for a single companion model at the best fit contrast for each epoch. The position angle of the best fit companion changes between 2010 and 2013. However, it does not change by the amount expected for a companion at $\sim 9.5$ AU in the outer disk plane. Figure \ref{fig:tcha_compfits} shows this; the $\chi^2$ minimum does not lie at the predicted position of an orbiting companion (hollow circles) for any of the datasets except for NaCo 2011. The apparent changes in position angle cannot be explained by a companion on a Keplerian orbit in the outer disk plane. A companion on either a very eccentric or very misaligned orbit can reproduce some of the observations, but this does not seem physically likely \cite{2015ApJ...801...85S}.

Figure \ref{fig:tcha_ims} shows the reconstructed images for each epoch of T Cha observations. The three-lobed structure in most of the reconstructed images suggests that a single companion model is too simple to adequately describe the data. The reconstructed images show that the best fit position angle of a single companion model fit almost always overlaps with the brightest lobe in the reconstructed image (white lines in Figure \ref{fig:tcha_ims}). A three companion model could describe a single epoch of observations if one point source were located at each lobe in the reconstructed images. While this model could better reproduce the data, the lack of orbital motion in the reconstructed images argues against it; three stationary companions would be unphysical. The reconstructed images suggest that the changing single companion position angle is caused by fitting an over-simplified model to the data, a multiple companion model would not make physical sense, and that a stationary model would provide a better explanation for the observations.

\subsection{Scattered Light Scenario}

\begin{figure} [ht]
\begin{center}
\includegraphics[height = 10cm]{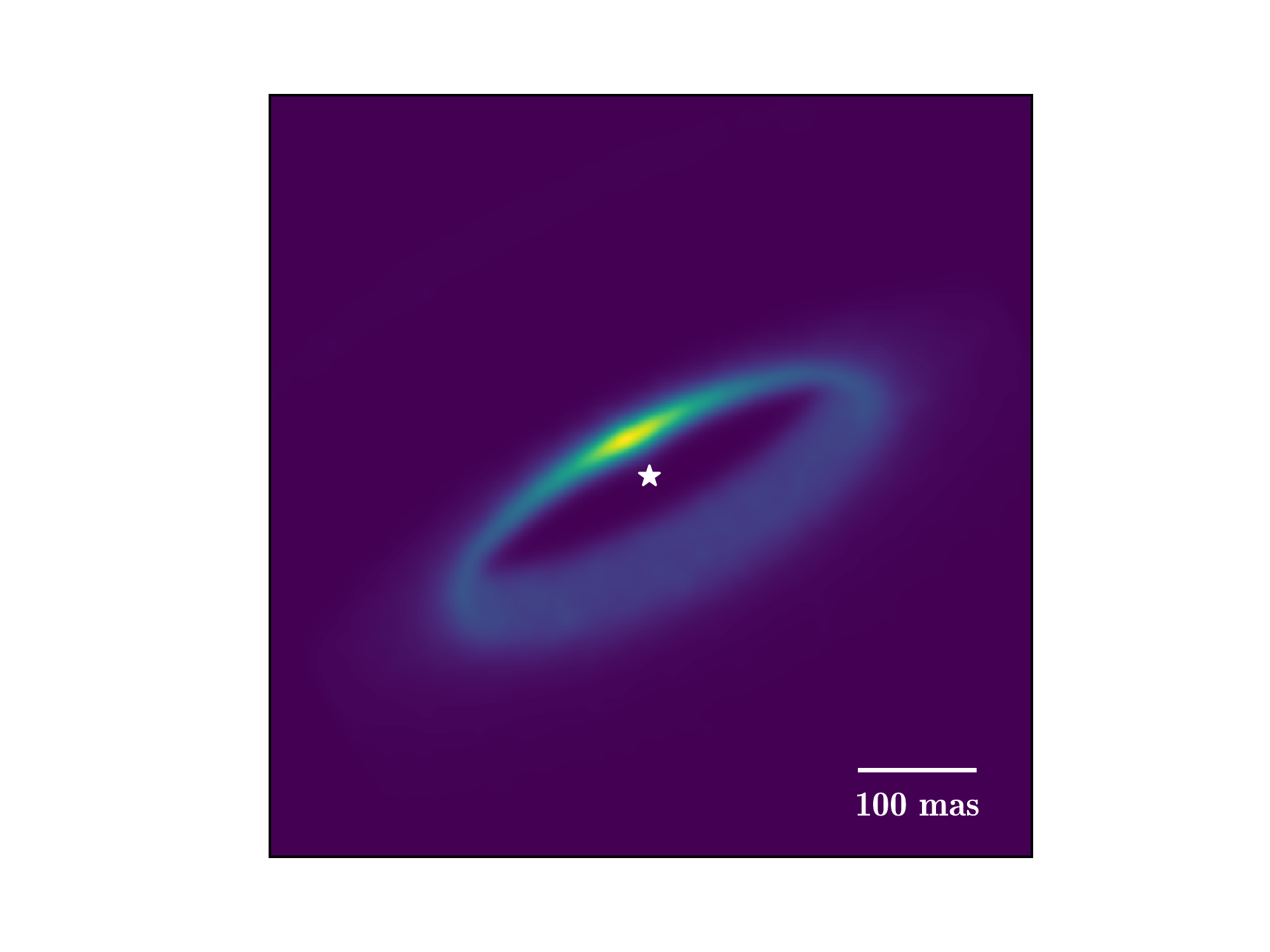}
\end{center}
\caption[tcha_inscat] 
{\label{fig:tcha_inscat} 
\emph{Hyperion} raytraced image of T Cha's transition disk using the disk parameters published in Huelamo et al. 2015.}
\end{figure}

We raytraced images of T Cha's outer disk to test whether the structure in the reconstructed images could be caused by scattered light. We used the open source radiative transfer code \emph{Hyperion}\cite{2011A&A...536A..79R}, and disk parameters found in Huelamo et al. 2015 \cite{2015A&A...575L...5H}. Figure \ref{fig:tcha_inscat} shows the L band raytraced disk image that we sampled like the masking observations. We simulated NRM datasets by sampling the raytraced images with the same (u,v) coverage and sky rotation as each epoch of T Cha observations. We also measured the scatter in the calibrated closure phases and squared visibilities for each T Cha dataset and added equal amounts of Gaussian noise to the simulated data. We then reconstructed images using both the BSMEM\cite{1994IAUS..158...91B} and MACIM\cite{2006SPIE.6268E..1TI} algorithms. The two algorithms produced comparable results. 

\begin{figure} [ht]
\begin{center}
\includegraphics[height = 10cm]{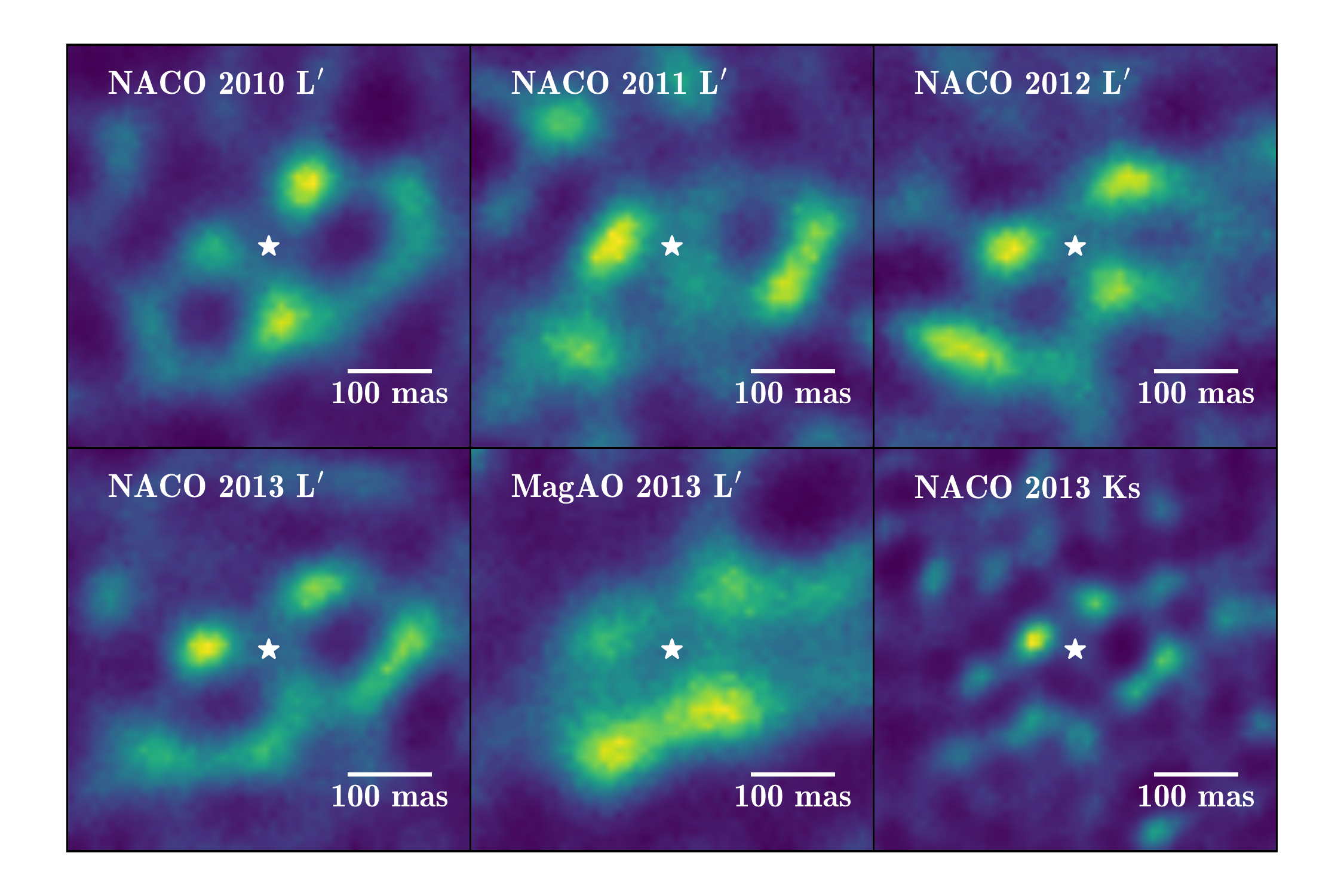}
\end{center}
\caption[tcha_uvcov_n] 
{\label{fig:tcha_uvcov_n} 
Reconstructed images from simulated observations of the disk shown in Figure \ref{fig:tcha_inscat}. For each simulated dataset we added Gaussian noise at the level of the actual observations and used identical sky rotation and (u,v) coverage.}
\end{figure}

Figure \ref{fig:tcha_uvcov_n} shows the reconstructed images for a single Gaussian noise realization added to each outer disk observation.     The images shown in the first four panels of Figure \ref{fig:tcha_uvcov_n} were reconstructed from observations using the NaCo mask at L$'$, with comparable sky rotation\cite{2015ApJ...801...85S}. Comparison of these panels shows that multi-epoch observations with the same mask and wavelength and similar sky rotation will produce nearly identical images. The same three-lobed structure is present in all four NaCo L$'$ images, and noise fluctuations cause flux to move between the lobes in different noise realizations.

Different masks at the same wavelength may yield different reconstructions for observations of the same source as well. The different (u,v) coverage and sky rotation in the MagAO 2013 L$'$ compared to NaCo 2010-2013 L$'$ observations produces different images.  Furthermore, different wavelength observations using the same mask may also result in different reconstructed images (NaCo 2013 Ks compared to NaCo 2010 - 2013 L$'$). This is because resolution scales with wavelength and thus different wavelength observations will have different beams.

The simulated disk observations can reproduce the structure in the reconstructed images shown in Figure \ref{fig:tcha_ims}. The changes in brightness for different observations and noise realizations are consistent with the brightness changes in images reconstructed from the actual data (compare NaCo 2010 and NaCo 2012 in Figure \ref{fig:tcha_ims}). Like the actual T Cha observations, fitting a single companion model to the datasets shown in Figure \ref{fig:tcha_uvcov_n} would result in a best fit position angle that jumps erratically between noise spikes. This suggests that the T Cha reconstructed images and single companion model fits are not caused by the presence of companions, but rather by noisy observations of scattered light from the outer disk.

\section{LkCa 15}
\subsection{Previously Published Observations}

\begin{figure} [ht]
\begin{center}
\includegraphics[height = 5cm]{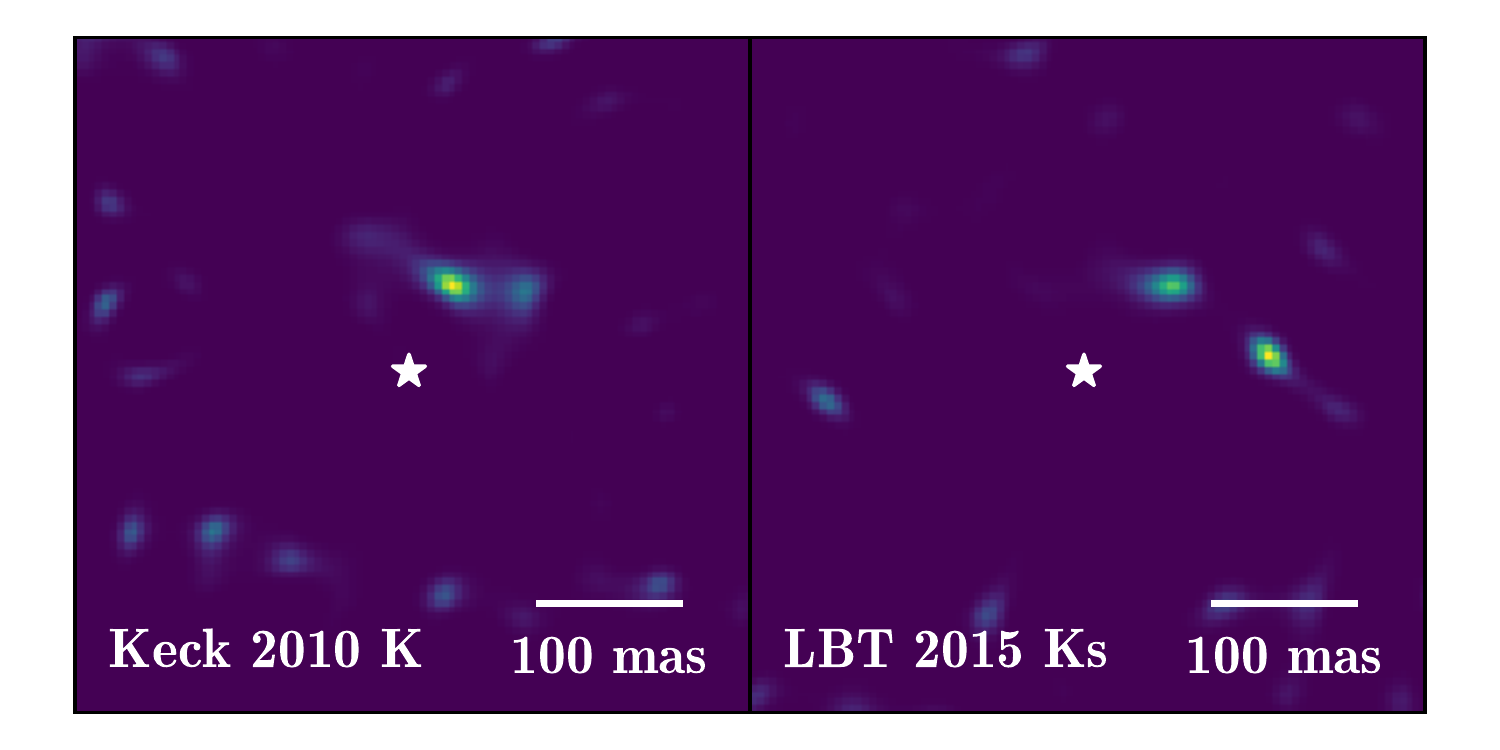}
\end{center}
\caption[lkca15_K] 
{\label{fig:lkca15_K} 
Reconstructed images from Keck (left) and LBT (right) K band NRM observations of LkCa 15.}
\end{figure}

In Sallum et al. 2015b\cite{2015Natur.527..342S}, we presented reanalyzed archival Keck infrared observations of LkCa 15\cite{2012ApJ...745....5K} as well as new LBT/LMIRCam\cite{2008SPIE.7013E..28H,2012SPIE.8446E..4FL} infrared and Magellan/MagAO\cite{Morzinski:2014,Close:2012} H$\alpha$ datasets. We fit multiple companion models to the Keck and LBT masking observations and also reconstructed images using both BSMEM\cite{1994IAUS..158...91B} and MACIM\cite{2006SPIE.6268E..1TI}. Figures \ref{fig:lkca15_L} and \ref{fig:lkca15_K} show the reconstructed images from L and K bands, respectively. In all the Keck observations and in the LBT L$'$ dataset we detect three point sources within the disk clearing (LkCa 15 b, c, and d), two of which (LkCa 15 b and c) coincide with LBT Ks detections. We detect one of the infrared point sources (LkCa 15 b) at H$\alpha$, an accretion tracer\cite{2014ApJ...783L..17Z,2012A&A...548A..56R,1994ApJ...426..669H}.

\subsection{New LBT L$'$ Observations}

\begin{figure} [ht]
\begin{center}
\includegraphics[height = 10cm]{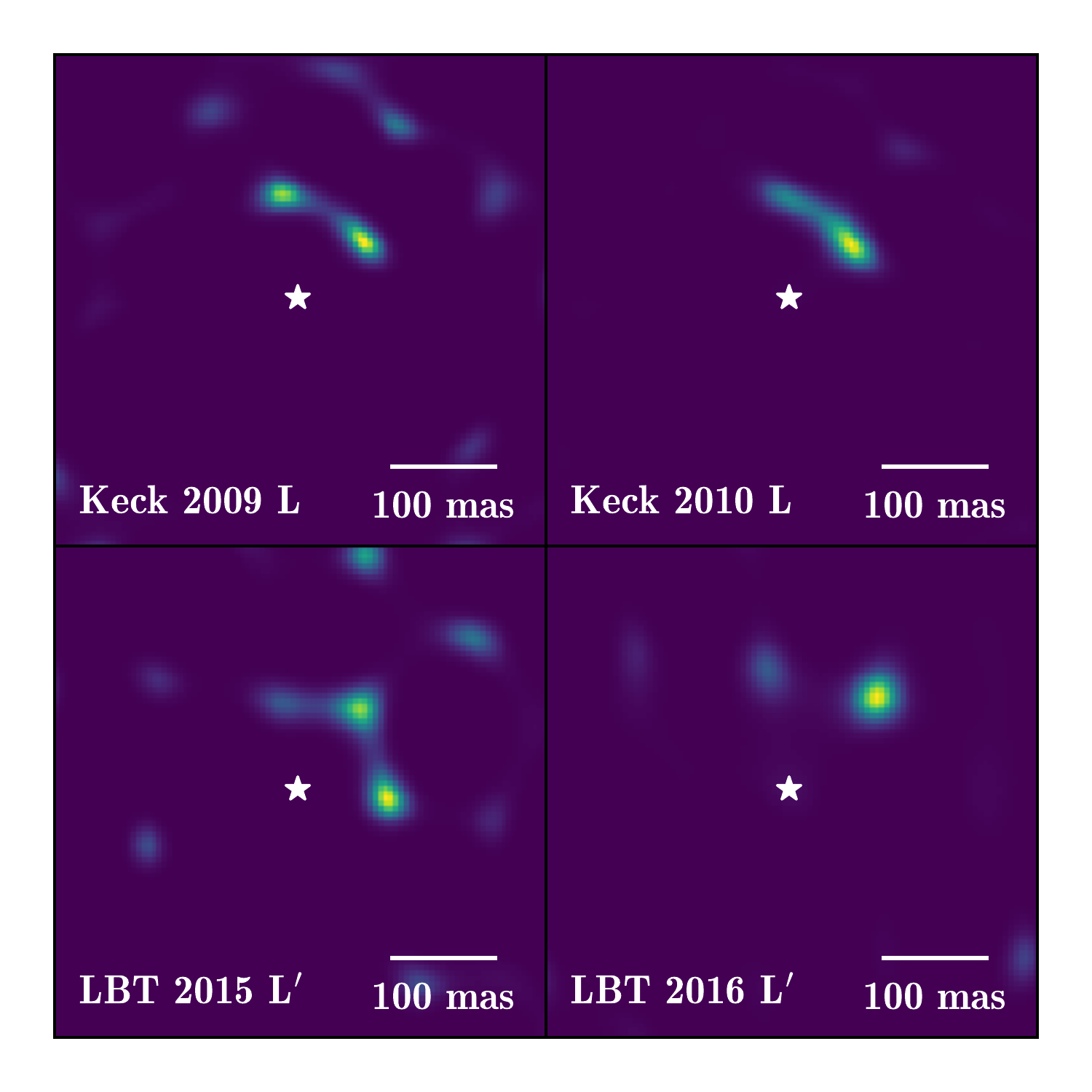}
\end{center}
\caption[lkca15_L] 
{\label{fig:lkca15_L} 
Reconstructed images from Keck (top row) and LBT (bottom row) L band NRM observations of LkCa 15. While LkCa 15 b is not detected in the new LBT observations, the noise levels and sky rotation present in the data are such that there is a $\sim43\%$ chance of a false non-detection. The new LBT observations are consistent with the three companion model.}
\end{figure}

In February 2016 we observed LkCa 15 at L$'$ using the LBT and LMIRCam's 12-hole mask. The scatter in these data was higher than those published in Sallum et al. 2015b (calibrated closure phase scatter of 2.5$^\circ$ compared to 1.9$^\circ$ in December 2014). The amount of sky rotation was also significantly lower (parallactic angles between 21$^\circ$ and 65$^\circ$ compared to -65$^\circ$ and 65$^\circ$ in December 2014). We fit multiple companion models and reconstructed images (see Figure \ref{fig:lkca15_L}) from these data. In both the model fitting and reconstructed images we detect two point sources consistent with the positions of LkCa 15 c and d.  

To test the significance of the 2016 LkCa 15 b data, we simulated datasets with identical (u,v) coverage and sky rotation angles as the new observations. We created a three-companion model with the separations and position angles set by the best fit orbit to the 2009-2015 masking observations. We fixed the contrast of each companion to the December 2014 L$'$ best fit. We added 1000 Gaussian noise realizations with standard deviation of $2.5^\circ$ to the simulated closure phases and reconstructed an image for each noise realization. For each reconstructed image we calculated the signal to noise ratio at the input position of LkCa 15 b by comparing the mean flux in an aperture around b to the average flux in all apertures not containing b, c, or d. LkCa 15 b's signal to noise was greater than 1 in $\sim57\%$ of noise realizations. Choosing a signal-to-noise ratio so low gives us the most conservative estimate on LkCa 15 b's false negative probability. Thus LkCa 15 b has at least a $\sim43\%$ chance of false non-detection; this is much higher than the false negative probability for LkCa 15 c ($\sim3\%$) due to b's closer separation and slightly lower flux. 

The new observations are consistent with the three companion scenario. While even low-quality data can potentially detect LkCa 15 c, we require better signal to noise than obtained in our 2016 observations to reliably detect b. To detect LkCa 15 b at the same confidence as c in the new dataset requires closure phases with scatter of $\sim1^\circ$.

\subsection{Multiple Companion Scenario}

\begin{figure} [ht]
\begin{center}
\includegraphics[height = 5cm]{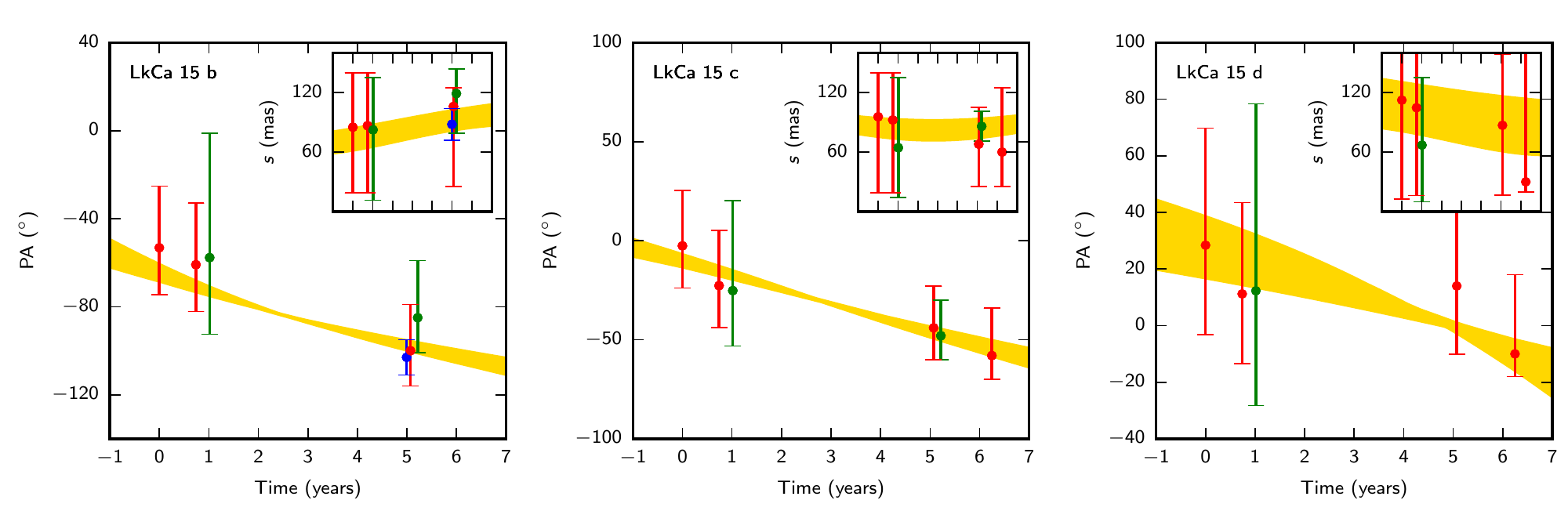}
\end{center}
\caption[lkca15_orbits] 
{\label{fig:lkca15_orbits} 
Position angle (large panels) and separation (insets) evolution over time for LkCa 15 b (left), c (center), and d (right). Red, green, and blue points show positions for sources detected at L$'$, Ks, and H$\alpha$, respectively, and error bars were generated using a $\chi^2$ interval on a grid. Yellow shading marks the 1 $\sigma$ allowed region from orbital fitting published in Sallum et al. 2015b. The last points in each of the two rightmost panels show positions for c and d from fits to the new LBT observations. These agree with the previously published orbital fits.}
\end{figure}

In LkCa 15 we observe smooth position angle changes of the best fit companion model over several years of observations. The positions between 2009 and 2016 agree with those expected for companions on distinct circular orbits in the outer disk plane. Figure \ref{fig:lkca15_orbits} shows the position evolution of the three companion model over the last $\sim6$ years, with the new LBT L$'$ positions of LkCa 15 c and d added. While the positions agree with those expected for Keplerian orbits, the uncertainties on the orbital separations are still quite large. For example, the uncertainties on LkCa 15 d's position are large enough that a constant position angle could go through all of the error bars. A longer time baseline and higher resolution observations would allow us to place better constraints on the point source positions and on their orbital parameters.

\subsection{Scattered Light Scenario}

LkCa 15's outer disk has been detected both at sub-mm and mm wavelengths\cite{2011ApJ...732...42A,2014ApJ...788..129I} and in the near infrared in scattered light\cite{2014A&A...566A..51T}. Since we have seen that scattered light can masquerade as a companion signal in NRM data, we simulated masking observations of LkCa 15's outer disk. We used \emph{Hyperion} to create ray traced images for the disk parameters and dust properties in Thalmann et al. 2014.\cite{2014ApJ...788..129I} The outer disk could not cause point sources within $\sim 100 - 200$ mas of the star in the reconstructed images. A more compact scattered light model is required to produce point sources at stellocentric distances consistent with those in the observed reconstructed images.

\begin{figure} [ht]
\begin{center}
\includegraphics[height = 10cm]{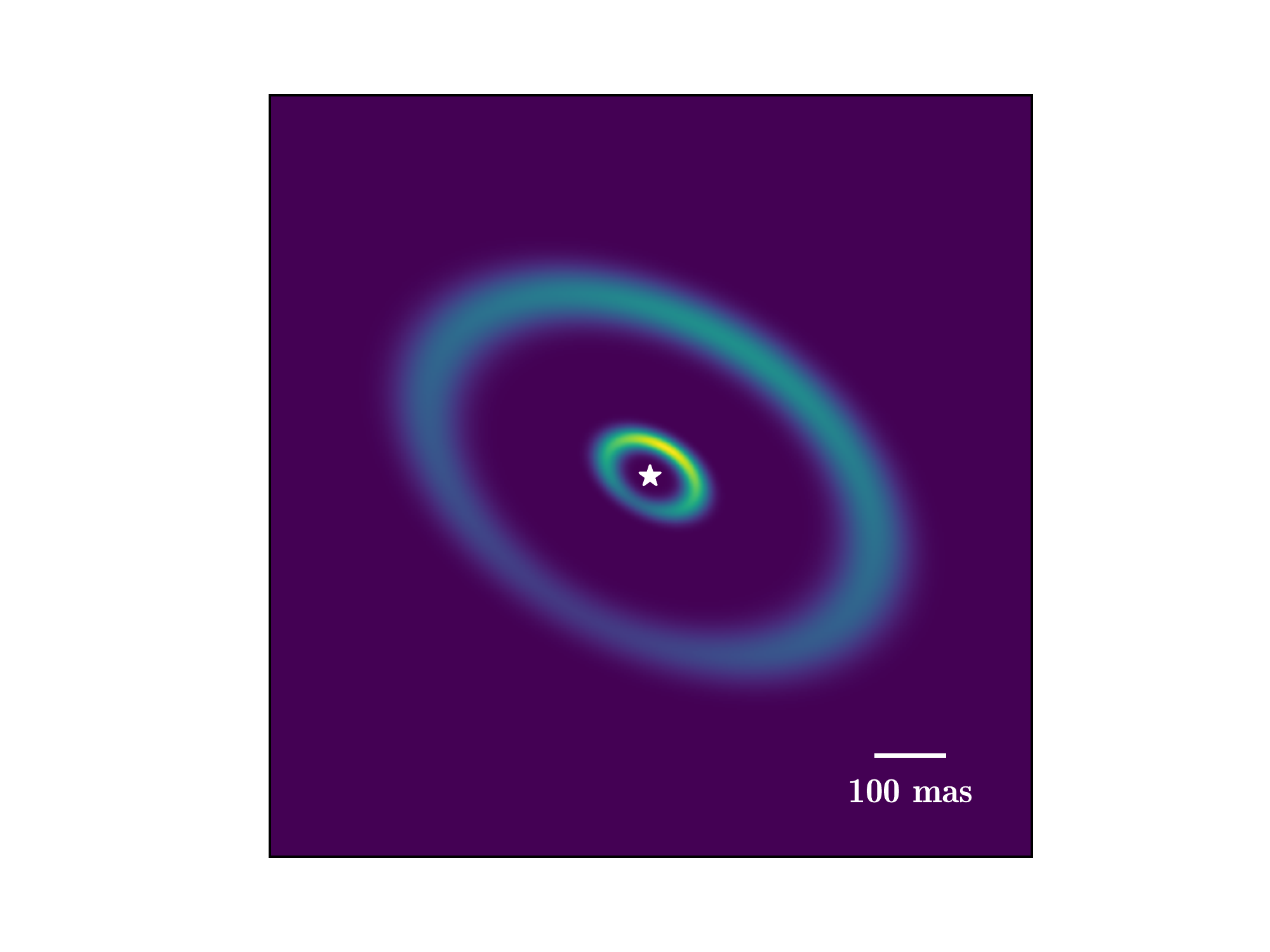}
\end{center}
\caption[lkca15_input] 
{\label{fig:lkca15_input} 
Input two disk model for LkCa 15 scattered light testing. This two disk model was motivated by recent observations of the inner disk and by the fact that simulated observations of the outer disk alone could not produce point sources close to the star in reconstructed images.}
\end{figure}

Recent SPHERE/ZIMPOL observations revealed an inner disk component in LkCa 15\cite{2015ApJ...808L..41T}. We simulated masking observations to test whether scattered light from a two disk model could cause low-separation point source signals in reconstructed images. Since the inner disk parameters are not well constrained, rather than raytrace images of both disks we used a parametric skewed-ring model.\cite{2009A&A...508..787K} We set the contrast of the outer disk to $\sim5\%$ to match previous scattered light observations\cite{2014A&A...566A..51T}. We then scaled the flux of the inner disk to be approximately twice that observed for the outer disk, to match optical polarimetric observations\cite{2015ApJ...808L..41T}. This model is shown in Figure \ref{fig:lkca15_input}. For each dataset, we used (u,v) coverage and sky rotation identical to the real observations, and added Gaussian noise at the level measured for the real data. 

\begin{figure} [ht]
\begin{center}
\includegraphics[height = 10cm]{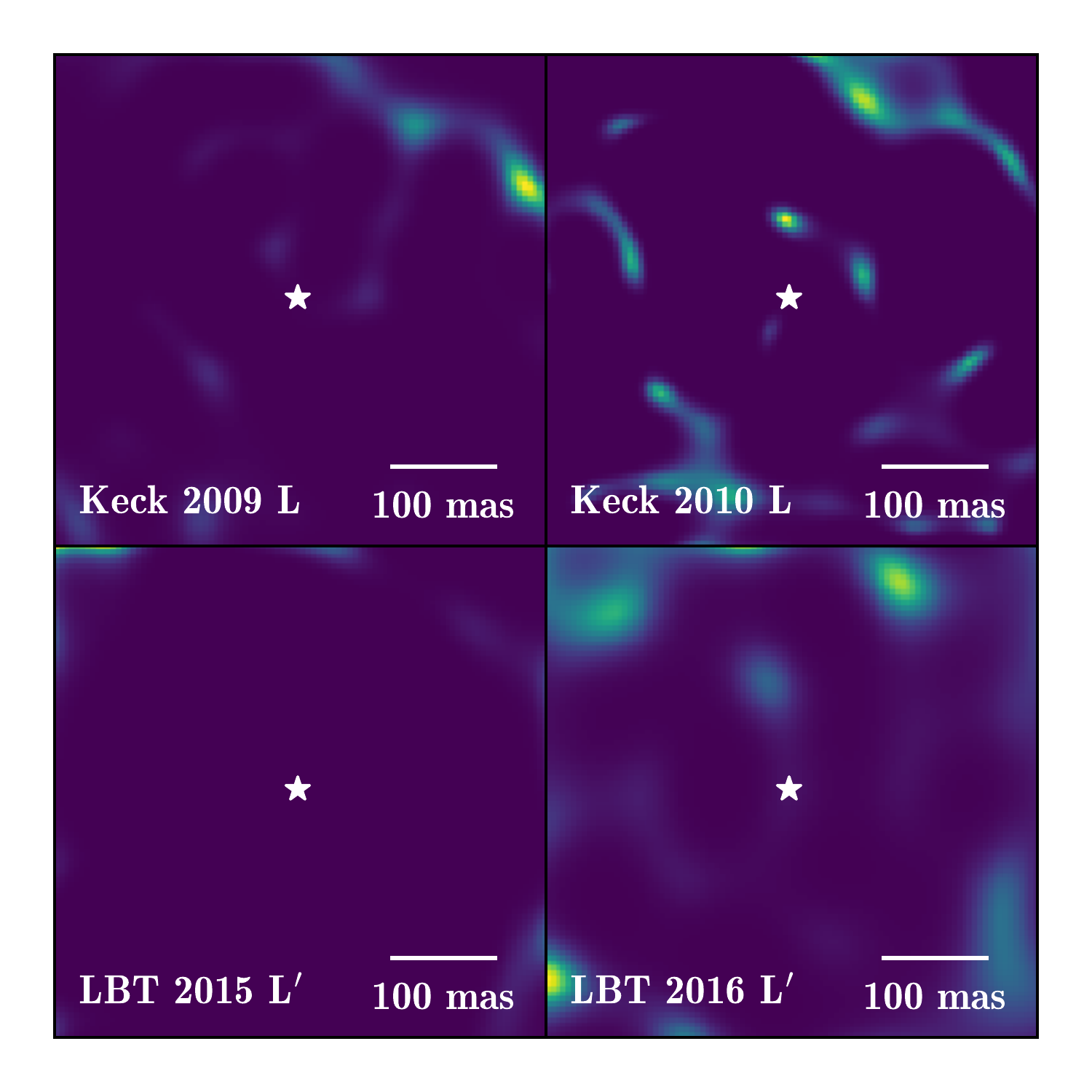}
\end{center}
\caption[lkca15_uvsim_L_n] 
{\label{fig:lkca15_uvsim_L_n} 
Reconstructed images for a single noise realization added to simulated L band observations of the LkCa 15 two disk model shown in Figure \ref{fig:lkca15_input}. At the noise levels present in the observations, scattered light from LkCa 15 does not always cause bright point sources close to the star.}
\end{figure}

\begin{figure} [ht]
\begin{center}
\includegraphics[height = 5cm]{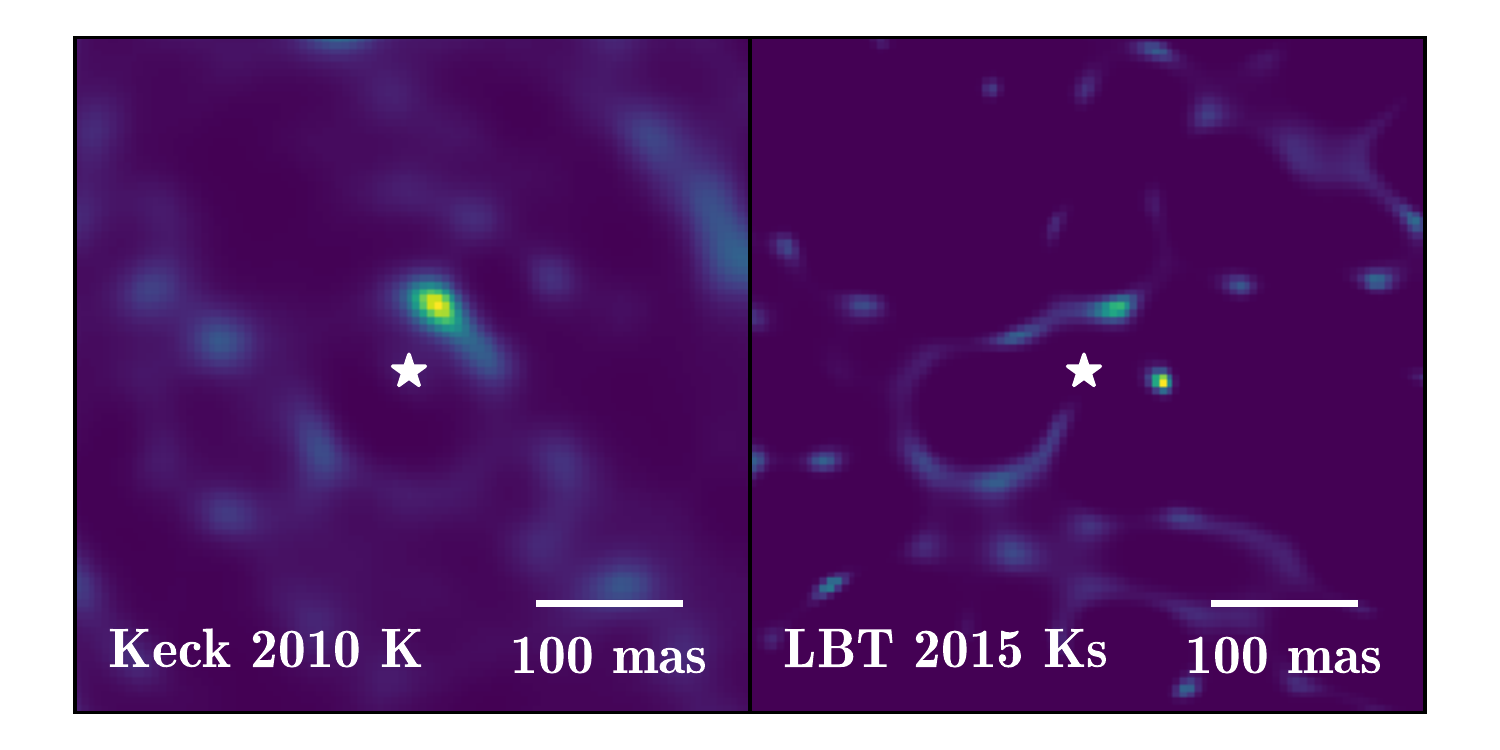}
\end{center}
\caption[lkca15_uvsim_K_n] 
{\label{fig:lkca15_uvsim_K_n} 
Reconstructed images for a single noise realization added to simulated K band observations of the two disk model shown in Figure \ref{fig:lkca15_input}.}
\end{figure}

Figures \ref{fig:lkca15_uvsim_L_n} and \ref{fig:lkca15_uvsim_K_n} show reconstructed images for the two disk model plus a single noise realization. In each panel, the sources within $\sim 100$ mas of the star to the northwest (upper right) are due to light from the inner disk, while the flux in that direction at larger separation is due to the outer disk. Comparing the top two panels of Figure \ref{fig:lkca15_uvsim_L_n} shows that adding noise can change the relative fluxes of the point sources caused by each disk. For example, flux at the position of the outer disk is much brighter in the 2009 Keck L band simulation compared to 2010, despite the fact that they have identical masks and wavelengths and similar sky rotation. Additionally, in the 2015 L$'$ simulation, the inner disk does not cause any bright point sources close to the star. While only a single noise realization is shown in each panel of Figure \ref{fig:lkca15_uvsim_L_n}, comparison of different noise realizations with the same scatter shows that noise spikes can cause flux to move between the inner and outer disk regions. These variations make it unlikely that scattered light from the inner disk would consistently cause point sources close to the star. This is inconsistent with the observations, which show close-in point sources in all epochs (see Figures \ref{fig:lkca15_L} and \ref{fig:lkca15_K}).

If scattered light from the inner disk caused point sources close to the star across multiple epochs, their positions would differ for observations made with different masks. Like the NaCo versus MagAO T Cha reconstructed images, the locations of point sources for simulated LBT disk observations are different from those for Keck. It would be extremely difficult for a scattering signal to cause consistent point source detections across multiple epochs using multiple masks.

We used the simulated disk observations to quantify the probability that scattered light would cause three point sources across multiple epochs.  To estimate the probability for a single observational epoch, we calculated the signal to noise at the positions of b, c, and d in 100 reconstructed images with different noise realizations. We measured the mean brightness in an aperture at the position of each companion and divided it by the mean flux of all apertures not containing b, c, or d. We also limited our noise measurement to separations interior to $\sim200$ mas, so that the average noise level did not include measurements of LkCa 15's outer disk flux. Out of 100 noise realizations, none resulted in detections at all three positions with signal to noise greater than 1. The probability that forward scattering from the inner disk caused the companion signals for a single epoch is thus $<0.01$. Assuming the six observations are independent, the probability that we would detect three point sources across all epochs is vanishingly small. 

Even if scattered light from LkCa 15's inner disk were causing multiple point sources to appear in the reconstructed images, their positions would remain fixed for multi-epoch datasets with the same mask and roughly the same sky rotation. A single companion fit to these data may show an apparent change in position angle as flux moves between each of the sources caused by the inner disk. A multiple companion fit to the scattering signal (with as many companions as point sources caused by the inner disk) would show no changes in position angle. An orbiting companion would cause a signal with a smoothly changing position angle that could not be caused by scattered light.

The sources seen in both the Keck and LBT reconstructed images have position angles that change with time. Comparing the top two panels of Figure \ref{fig:lkca15_L}, the position angles of the sources change by several degrees between 2009 and 2010. The two sources seen in both bottom panels of Figure \ref{fig:lkca15_L} also have changing position angles. This agrees with the fact that previously published Keck K band images show three sources with position angles changing by $\sim4^\circ$ per year \cite{2014IAUS..299..199I}. A forward scattering scenario could not cause position angle changes like these.

The observed infrared fluxes argue against the scattered light scenario as well\cite{2015Natur.527..342S}. Since dust opacity increases with decreasing wavelength, a scattering signal will have higher flux at shorter wavelength. However the infrared sources detected in both Keck and LBT observations are brighter at L band than at K band \cite{2012ApJ...745....5K,2014IAUS..299..199I,2015Natur.527..342S}. A non-detection at Keck at H band\cite{2014IAUS..299..199I}, and a strong detection at M band\cite{2014IAUS..299..199I} argue further against scattered light as the cause of the structure in reconstructed images. Lastly, a scattered light model could not cause the H$\alpha$ detection, since dust opacity at H$\alpha$ and in the nearby continuum would be equal.

\section{Conclusions}

We presented multi-epoch NRM observations of the T Cha and LkCa 15 transition disks. For both objects we find best fit companion models with position angles that change with time. In the case of T Cha we showed that the data are better explained by scattered light from the outer disk than by an orbiting companion. While the position angle of the best fit companion does change, it does not do so in a Keplerian way or even monotonically. This apparent motion is most likely caused by fitting a single companion fit to noisy observations of an extended scattered light signal.

For LkCa 15 we show that at the noise levels present in the observations, scattered light from the inner disk can cause point sources close to the star in reconstructed images, but not consistently. Furthermore, scattering is highly unlikely to cause multiple point sources to appear at consistent positions for different masks, and even less likely to cause point sources with smoothly changing position angles. A non-detection at H band, a strong detection at M band, and the H$\alpha$ detection argue against the scattered light hypothesis as well. Unlike T Cha, the position angle changes observed for LkCa 15 agree with those for companions on Keplerian orbits aligned with the outer disk. For these reasons, we argue that the multiple companion scenario most naturally explains the NRM observations of LkCa 15. 

The tests presented here highlight the importance of using a combination of model fitting and image reconstruction in non-redundant masking observations. Reconstructed images can reveal a simple model such as a single companion to be inadequate, as in the case of T Cha. However, our ability to reconstruct an image of an extended source is limited by the (u,v) coverage, sky rotation, and noise levels of the observations. For this reason, it is useful to use modeling to check whether extended sources such as disks could create point source structure in reconstructed images. NRM is a powerful technique for detecting close in companions, but care must be taken to distinguish between true companion signals and extended sources. 

The ability to distinguish between companions and scattered light improves with increased (u,v) coverage (and thus larger amounts of recoverable phase information) as well as with increased resolution. For example, simulated images using the co-phased LBTI's 12 hole mask easily recover a single extended source for both T Cha's disk and LkCa 15's outer disk with the same amount of sky rotation as the LBT L$'$ LkCa 15 data. A close-in LkCa 15 inner disk also appears much more extended in dual-aperture LBTI masking data compared to the intra-aperture observations shown here. Future observations of transition disks using the co-phased LBTI will allow for unambiguous detection of companions at even closer separations than those seen in LkCa 15. 

\acknowledgments 
This work was supported by NSF AAG grant no. 1211329 and NASA OSS grant NNX14AD20G. This material is based upon work supported by the National Science Foundation under grant no. 1228509. This work was performed in part under contract with the California Institute of Technology (Caltech) funded by NASA through the Sagan Fellowship Program executed by the NASA Exoplanet Science Institute. This material is based upon work supported by the National Science Foundation Graduate Research Fellowship under grant no. DGE-1143953. Any opinion, findings, and conclusions or recommendations expressed in this material are those of the authors(s) and do not necessarily reflect the views of the National Science Foundation. 

\bibliography{arxiv} 
\bibliographystyle{spiebib} 

\end{document}